\documentclass[aps,prl,two column,showpacs,superscriptaddress,10pt]{revtex4-2}

\usepackage{amsmath, amssymb, bbm}
\usepackage{graphicx}
\usepackage{fancybox}
\usepackage{mathrsfs}
\usepackage{color}
\usepackage{booktabs}
\usepackage{placeins}
\usepackage[linktocpage=true]{hyperref}
\makeatletter
\newcommand*{\bigs}[1]{{\hbox{$\left#1\vbox to9.0\p@{}\right.\n@space$}}}
\makeatother % notwendig beim Laden von breqn
\usepackage{breqn}
\usepackage[caption=false]{subfig}
\usepackage{array}
\usepackage{multirow}

\newcommand{\de}{\mathrm{d}}

\renewcommand{\dag}{^\dagger}
\renewcommand{\vec}[1]{\boldsymbol #1}

\renewcommand{\i}{\mathrm{i}}
\newcommand{\h}{\hspace{1pt}}
\newcommand{\mh}{\hspace{-1pt}}
\newcommand{\hh}{\hspace{0.5pt}}
\newcommand{\mhh}{\hspace{-0.5pt}}
\newcommand{\e}{\mathrm e}

\newcolumntype{C}[1]{>{\centering\arraybackslash}m{#1}}

\DeclareOldFontCommand{\rm}{\normalfont\rmfamily}{\mathrm}

\usepackage[T1]{fontenc}
\newcommand{\db}{\textnormal{\dj}}

\begin{document}

\title{Triplet superconductivity in the Rashba$^4$ model}

\author{G.\,A.\,H.~Schober}
\email{g.schober@thphys.uni-heidelberg.de}
\affiliation{Institute for Theoretical Physics\char`,{}\,Heidelberg University\char`,{}\,Philosophenweg 19\char`,{}\,69120 Heidelberg\char`,{}\,Germany}
\affiliation{Institute for Theoretical Solid State Physics\char`,{}\,RWTH Aachen University\char`,{}\,52074 Aachen\char`,{}\,Germany}

\author{M.~Salmhofer}
\email{salmhofer@uni-heidelberg.de}
\affiliation{Institute for Theoretical Physics\char`,{}\,Heidelberg University\char`,{}\,Philosophenweg 19\char`,{}\,69120 Heidelberg\char`,{}\,Germany}

\date{\today}

\begin{abstract}
The Rashba$^4$ model is a lattice model of interacting fermions with a tight-binding dispersion relation that exhibits four Rashba band crossing points at the same energy. The fermions interact weakly, by an on-site repulsion and an additional nearest-neighbour two-body interaction, which may be attractive or repulsive. We study this model when the Fermi energy is at or close to the band crossings, by a combination of renormalization group flow equations and mean-field theory. We find that the scattering processes between different Rashba points drive superconductivity of an unusual triplet type not found in models with a single Rashba point. 
\end{abstract}

%\pacs{75.20.Ck, 71.70.Ej, 85.75.-d, 31.15.A-, 71.15.-m}
% PACS, the Physics and Astronomy classification Scheme.
% \keywords{Suggested keywords}
% Use showkeys class option if keyword display desired

\maketitle
%\tableofcontents
%
%\newpage

% \section{Rashba$^4$ model}

% Rashba paper Eq. (A123)

Fermionic models with a Rashba dispersion relation and a short-range two-body interaction exhibit a variety of competing correlations and potential ordered phases, in particular superconductivity \cite{vafek2011,Schober}. In this paper and a more detailed companion paper \cite{SchoSa2}, we consider a general class of tight-binding models where several Rashba points arise naturally. We use renormalization group (RG) methods to derive a scale-dependent effective action. The existence of several Rashba points at the Fermi level leads to new interaction terms in this effective action. We study the interplay of these interactions in the framework of the g'ology approximation, where the focus is on the most singular contributions, i.e., the coupling function is reduced to a finite number of terms that describe the scattering of particles from one Rashba point to another in the Brillouin zone. In the case where the dominant instability is of superconducting type, we use the RG effective action as an input for BCS-type mean-field theory, by which we calculate the gap function and the order parameter in the symmetry-broken phase.

%
%older intro
%
%Much work has been done on competing correlations and potential ordered phases, in particular superconductivity, in fermionic models with a Rashba dispersion relation \cite{vafek2011,Schober}. 
%In this paper and a more detailed companion paper \cite{SchoSa2}, we consider a general class of tight-binding models where several Rashba points arise naturally. %We study these models using both renormalization group (RG) methods and BCS-type mean field theory. The existence of several Rashba points at the Fermi level leads to new interaction terms in the flowing effective action RG. We study the interplay of these interactions in the framework of the g'ology approximation, where the focus is on the most singular contributions, i.e., the coupling function is reduced to a finite number of terms that describe scattering of particles from one Rashba point to another in the Brillouin zone. 

\begin{figure}[t]
\subfloat[Energy dispersion\label{Dispersion}]{
 \includegraphics[width=0.54\columnwidth]{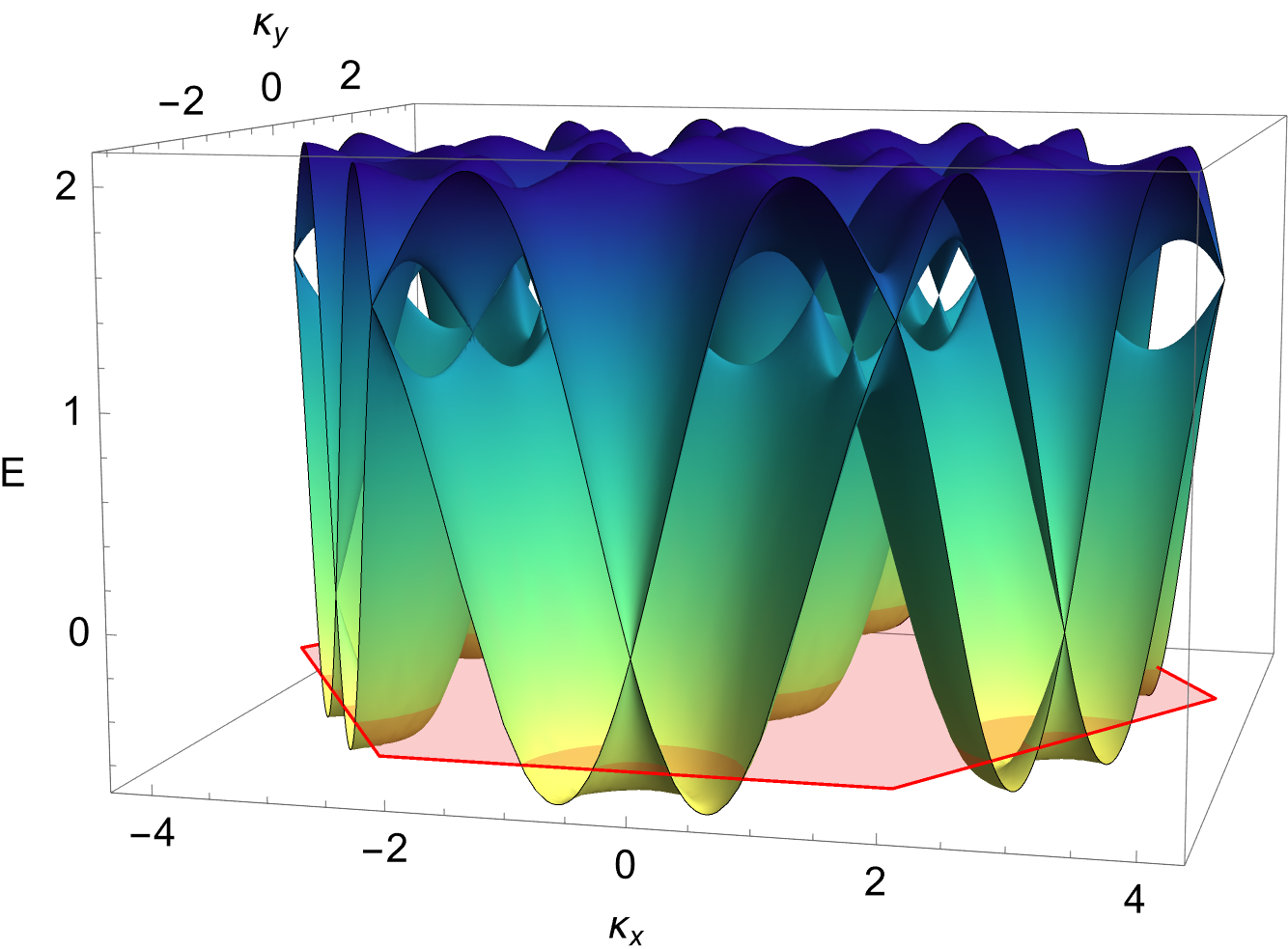}}
\hfill
\subfloat[Fermi lines\label{Fermi}]{
 \raisebox{0.2cm}{\includegraphics[width=0.4\columnwidth]{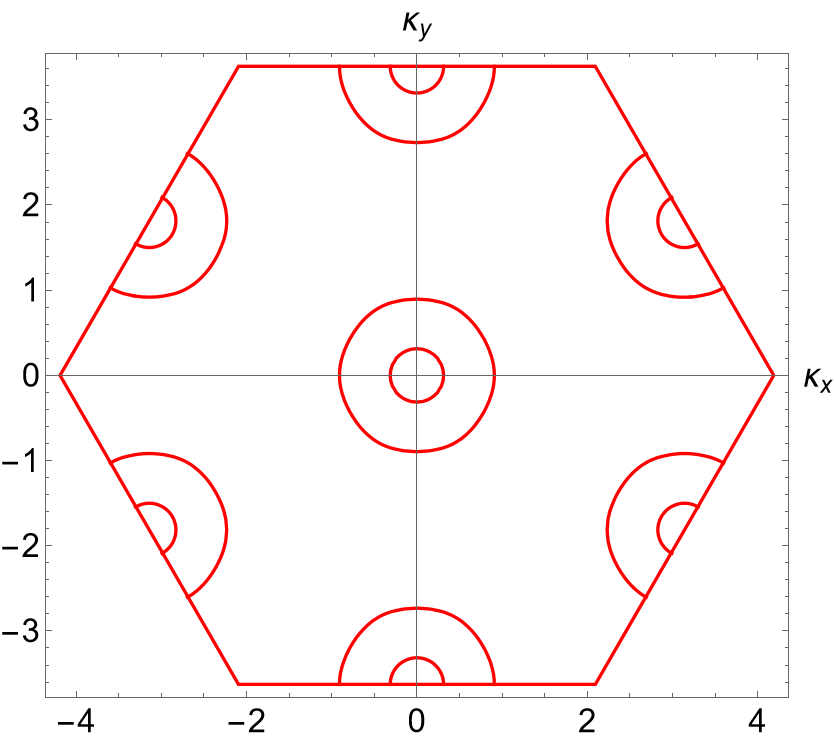}}}
 \caption{Rashba$^4$ model for parameters $\alpha = 2$ and $\gamma = 1$, with Fermi energy below the band crossings. \label{fig_disp_5}} \medskip
\end{figure}

The Rashba$^4$ model is a tight-binding model defined on the two-dimensional hexagonal lattice by the Hamiltonian
\begin{equation}
\hat H^0 = \sum_{\vec R, \h \vec R'} \h \sum_{s, \h s'} H^0_{ss'}(\vec R - \vec R') \, \hat a_s\dag(\vec R) \h \hat a_{s'}(\vec R') \,,
\end{equation}
where $\vec R$ , $ \vec R'$ are direct lattice vectors and $s$, $s'$ are spin indices.
% g-ology (2.1)
The $(2 \times 2)$ matrix $H^0$ can be expanded in terms of the Pauli matrices $\vec \sigma = (\sigma_x, \sigma_y, \sigma_z)^{\mathrm T}$, i.e.,
\begin{align} \label{gen_H}
 H^0(\vec R) = f(\vec R) + \vec g(\vec R) \cdot \vec \sigma \,.
\end{align}
The model includes next-nearest neighbor hopping to the vector $2 \vec R_1$, where $\vec R_1$ is a primitive vector of the hexagonal lattice, which by our convention points in the positive $x$ direction. Concretely, we define
\begin{align} \label{def_model}
 f(2\hh \vec R_1) &  := -\gamma/6 \,, \\[5pt] g_y(2 \hh \vec R_1) & := \i \hh \alpha/6 \,,
\end{align}
with parameters $\alpha, \gamma \in \mathbb R$, and we set $g_z(\vec R) \equiv 0$.
Furthermore, we assume Hermiticity, time-reversal symmetry, three-fold rotation symmetry $C_3$ about the $z$ axis, and mirror reflection symmetry $M_x$ with respect to the $x$-axis \cite{Schober}. As a result of these conditions, one can show that in dual space,
% g-ology (2.65)
\begin{align}
 f(\vec k) & = -\frac \gamma 3 \, \Big( \cos(\vec k \cdot 2 \hh \vec R_1) \\[3pt] \nonumber
&  \quad \, + \cos(\vec k \cdot C_3 \hh (2 \hh \vec R_1)) + \cos(\vec k \cdot C_3^{-1} (2 \hh \vec R_1)) \Big) \,, \label{zw_2} \\[5pt]
g_x(\vec k) & = \frac{\alpha \h \sqrt 3}{6}\, \Big( {-\sin(\vec k \cdot C_3 \h (2 \hh \vec R_1))} \\[3pt] \nonumber
&  \quad \, +  \sin(\vec k \cdot C_3^{-1} (2 \hh \vec R_1)) \Big) \\[5pt]
g_y(\vec k) & = \frac \alpha 6 \, \Big( 2 \h \sin(\vec k \cdot 2 \hh \vec R_1)  \\[3pt]  \nonumber
&  \quad \, - \sin(\vec k \cdot C_3 \h (2 \hh \vec R_1)) - \sin(\vec k \cdot C_3^{-1} (2 \hh \vec R_1)) \Big) \,.
\end{align}
% g-ology (2.86) ff.
The energy bands and Fermi lines of this model are shown in Fig.~\ref{fig_disp_5} (for Fermi energy $E_{\mathrm F} < 0$). Band crossings appear at each time-reversal invariant momentum $\vec k$ (where $\vec k = -\vec k$), i.e., at the center $O = (0, 0)^{\mathrm T}$ of the Brilllouin zone and at the three inequivalent boundary points
\begin{align}
 A = \frac{2\pi}{a_0} \begin{pmatrix} 0 \\[5pt] \frac{1}{\sqrt 3} \end{pmatrix} , \ B =\frac{\pi}{a_0}  \begin{pmatrix} -1 \\[5pt] \frac{1}{\sqrt 3} \end{pmatrix} , \ C = \frac{\pi}{a_0} \begin{pmatrix} -1\\[5pt] -\frac{1}{\sqrt 3} \end{pmatrix}
\end{align}
For each $X \in \{O, A, B, C\}$ and for $\vec \kappa = a_0 \hh \vec k = (\kappa_x, \kappa_y)^{\mathrm T}$, where $a_0 = \lvert \vec R_1 \rvert$ denotes the lattice constant, the Hamiltonian is  given to second order in $\vec \kappa$ by
\begin{align} \label{RH}
 H^0(X + \vec \kappa) = \gamma \h \lvert \vec \kappa \rvert ^2 + \alpha \h (\kappa_x \hh \sigma_y - \kappa_y \hh \sigma_x) \,,
\end{align}
which is the Rashba Hamiltonian. Hence, the model is approximated by the Rashba Hamiltonian near each of the four time-reversal invariant momenta $\{O, A, B, C\}$, which is why we call it the Rashba$^4$ model. The general case is treated in \cite{SchoSa2}. %The eigenvalues of this Hamiltonian are given by

% \section{Interaction parameters}

To the free Hamiltonian defined above, we add an interaction term, $\hat H = \hat H^0 + \hat V$, which consists of an onsite and a nearest-neighbor density-density interaction,
\begin{align} \label{ini_int}
 \hat V =2 \h U_0 \, \sum_{\vec R} : \mh \hat n_{\uparrow}(\vec R) \, \hat n_{\downarrow}(\vec R) \mh : \h + \ \frac{U_1}{2} \mh \sum_{\langle \vec R, \h \vec R' \rangle} \hat n(\vec R) \, \hat n(\vec R') \,.
\end{align}
Here, $\hat n_s(\vec R) = \hat a_{s}^{\dagger}(\vec R) \h \hat a_s(\vec R)$ is the spin-resolved density operator at lattice site $\vec R$, $\hat n(\vec R) = \hat n_{\uparrow}(\vec R) + \hat n_{\downarrow}(\vec R)$, and the summation in the second term is over all nearest-neighbor vectors $\vec R$ and $\vec R'$. We further assume that the first term is repulsive, i.e., $U_0 > 0$, while the second term may be of any sign.

\begin{table}[t]
\small
\renewcommand{\arraystretch}{1.2}
\begin{tabular}{p{0.24\columnwidth}p{0.54\columnwidth}p{0.15\columnwidth}}
\toprule[1pt]
Graphical & Coupling & Initial \\
\midrule[1pt]
\raisebox{\ht\strutbox-\height}{\includegraphics[height=1.5cm,angle=90,origin=c]{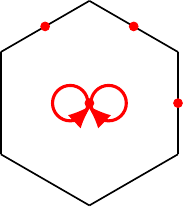}} & $a_1 = \mathrm{Re}\ V_{+-+-}(O,O,O,O)$ & $U_0 + 3 \, U_1$ \\
\midrule[1pt]
\raisebox{\ht\strutbox-\height}{\hspace{-0.1cm} \includegraphics[height=1.5cm,angle=90,origin=c]{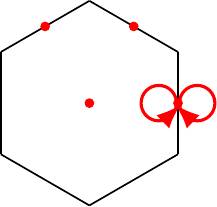} } \vspace{0.1cm} & $b_1 = \mathrm{Re}\ V_{+-+-}(A,A,A,A)$ & $U_0 + 3 \, U_1$ \\
\midrule[1pt]
\raisebox{\ht\strutbox-\height}{\hspace{-0.1cm} \includegraphics[height=1.5cm,angle=90,origin=c]{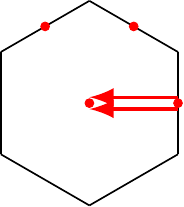} } & $c_1 = \mathrm{Re}\ V_{+-+-}(O,O,A,A)$ & $U_0 - U_1$ \\
\midrule[1pt]
\raisebox{\ht\strutbox-\height}{\hspace{-0.1cm} \includegraphics[height=1.5cm,angle=90,origin=c]{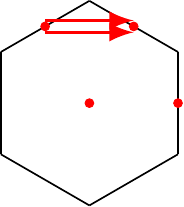} } & $d_1 = \mathrm{Re}\ V_{+-+-}(B,B,C,C)$ & $U_0 - U_1$ \\
\midrule[1pt]
\multirow{6}{*}{\includegraphics[height=1.5cm,angle=90,origin=c]{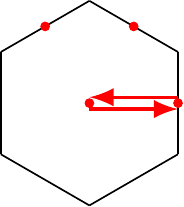}} & $e_1 = \mathrm{Re}\ V_{+ + + +}(O, A, O, A)$ & $4 \, U_1$ \\
& $e_2 = \mathrm{Im}\ V_{+ + + -}(O, A, O, A)$ & $0$ \\
& $e_3 = \mathrm{Im}\ V_{+ + - +}(O, A, O, A)$ & $0$ \\
& $e_4 = \mathrm{Re}\ V_{+ + - -}(O, A, O, A)$ & $0$ \\
& $e_5 = \mathrm{Re}\ V_{+ - + -}(O, A, O, A)$ & $U_0 + 3 \h U_1$ \\
& $e_6 = \mathrm{Re}\ V_{+ - - +}(O, A, O, A)$ & $-U_0 + U_1$ \\
\midrule[1pt]
\multirow{7}{*}{\includegraphics[height=1.5cm,angle=90,origin=c]{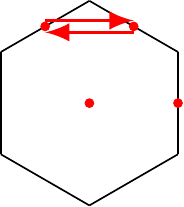}} & $f_1 = \mathrm{Re}\ V_{+ + + +}(B, C, B, C)$ & $4 \, U_1$ \\
& $f_2 = \mathrm{Re}\ V_{+ + + -}(B, C, B, C)$ & $0$ \\
& $f_3 = \mathrm{Im}\ V_{+ + + -}(B, C, B, C)$ & $0$ \\
& $f_4 = \mathrm{Re}\ V_{+ + - -}(B, C, B, C)$ & $0$ \\
& $f_5 = \mathrm{Re}\ V_{+ - + -}(B, C, B, C)$ & $U_0 + 3 \, U_1$ \\
& $f_6 = \mathrm{Re}\ V_{+ - - +}(B, C, B, C)$ & $-U_0 + U_1$ \\
& $f_7 = \mathrm{Im}\ V_{+ - - +}(B, C, B, C)$ & $0$ \\
\midrule[1pt]
\multirow{8}{*}{\includegraphics[height=1.5cm,angle=90,origin=c]{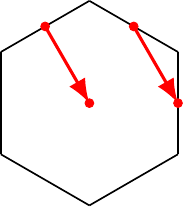}} & $h_1 = \mathrm{Im}\ V_{+ + + +}(O, A, B, C)$ & $0$ \\
& $h_2 = \mathrm{Re}\ V_{+ + + -}(O, A, B, C)$ & $0$ \\
& $h_3 = \mathrm{Im}\ V_{+ + + -}(O, A, B, C)$ & $0$ \\
& $h_4 = \mathrm{Im}\ V_{+ + - -}(O, A, B, C)$ & $0$ \\
& $h_5 = \mathrm{Re}\ V_{+ - + +}(O, A, B, C)$ & $0$ \\
& $h_6 = \mathrm{Re}\ V_{+ - + -}(O, A, B, C)$ & $U_0 - U_1$ \\
& $h_7 = \mathrm{Im}\ V_{+ - + -}(O, A, B, C)$ & $0$ \\
& $h_8 = \mathrm{Re}\ V_{+ - - -}(O, A, B, C)$ & $0$ \\
\bottomrule[1pt]
\end{tabular}
\caption{Real couplings and their initial conditions. For better readability, we write spin indices as $\uparrow \, = +$ and $\downarrow \, = -$\h. \label{tab_rename}}
\end{table}

To set up the RG flow, we consider an effective interaction kernel $V^\Lambda_{s_1 s_2 s_3 s_4}(\vec k_1, \vec k_2, \vec k_3, \vec k_4 = \vec k_1 + \vec k_2 - \vec k_3)$ depending on the RG scale $\Lambda$, which corresponds to the interaction operator \eqref{ini_int} at the initial scale $\Lambda = \Lambda_0$. Furthermore, we project each  momentum argument to the respective closest momentum in the set $\{O, A, B, C \}$  of the four time-reversal invariant momenta. Hence, with each spin index taking values in $\{\uparrow, \downarrow\}$, there are in principle $4^3 \times 2^4 = 1024$ complex interaction parameters. However, assuming that the effective interaction has the same symmetries as the free Hamiltonian, i.e., Hermiticity, time-reversal symmetry, three-fold rotation symmetry $C^3$, and mirror reflection symmetry $M_x$, one can show that several of these couplings coincide, and there are in fact only 25 independent {\itshape real} interaction parameters. These are listed in Table \ref{tab_rename}, which also provides a graphical representation and the initial condition (at $\Lambda = \Lambda_0$) for each interaction parameter.

% \section{G'ology flow}

From the RG equation for the effective interaction kernel \cite{SH01, Schober}, one can derive the {\itshape g'ology equations,} i.e., the 25 coupled flow equations for the real couplings listed in Table \ref{tab_rename}. Several couplings have identical initial conditions, and our numerical solution shows that some couplings even remain degenerate in the flow. Concretely, we find that 
$a_1 = b_1$, $c_1 = d_1 = h_6$, $e_1 = f_1$, $e_5 = f_5$, and $e_6 = f_6$, while all other couplings vanish identically. 
Thus, we obtain the following simplified RG equations, 
i.e., a set of five coupled differential equations for the real couplings $a_1$, $c_1$, $e_1$, $e_5$, and $e_6$:

\newpage
\begin{dgroup*} 
\begin{dmath*} \label{simpl_diff_first}
a_1'=
-2 \h I^-(-,+) \, (a_1^2+3c_1^2) \\ 
+3 \h I^+(-,-) \, (e_1^2+2 e_1 e_5+e_5^2-e_6^2) \\ 
+ I^+(-,+) \, (-2a_1^2-3 (e_1^2-2 e_1 e_5+e_5^2+e_6^2)) \\ 
\end{dmath*} 
\begin{dmath*} 
c_1'=
-4 \h I^-(-,+) \, c_1 (a_1+c_1) \\ 
-2 \h I^+(-,-) \, c_1 (e_1+e_5+e_6) \\ 
-2 \h I^+(-,+) \, c_1 (2 c_1+e_1+e_5-e_6) \\ 
\end{dmath*} 
\begin{dmath*} 
e_1'=
- I^-(-,-) \, e_1^2 \\ 
- I^-(-,+) \, e_1^2 \\ 
+ I^+(-,-) \, (-3 c_1^2+e_1^2+4 e_1 e_5+2e_5^2+2 a_1 (e_1+e_5)-2 e_1 e_6-e_6^2) \\ 
+ I^+(-,+) \, (-3 c_1^2-2a_1 e_1+e_1^2+2 a_1 e_5-4 e_1 e_5+2 e_5^2+2 e_1 e_6-e_6^2) \\ 
\end{dmath*} 
\begin{dmath*} 
e_5'=
- I^-(-,-) \, (e_5+e_6)^2 \\ 
+ I^-(-,+) \, (-4c_1^2-(e_5-e_6)^2) \\ 
+ I^+(-,-) \, (-3 c_1^2+2 e_1^2+4 e_1 e_5+e_5^2+2 a_1 (e_1+e_5)) \\ 
+ I^+(-,+) \, (-3 c_1^2-2 e_1^2+2 a_1(e_1-e_5)+4 e_1 e_5-3 e_5^2) \\ 
\end{dmath*} 
\begin{dmath*} \label{simpl_diff_last}
e_6'=
- I^-(-,-) \, (e_5+e_6)^2 \\ 
+ I^-(-,+) \, (4 c_1^2+(e_5-e_6)^2) \\ 
+ I^+(-,-) \, (-3 c_1^2-e_1^2-2 e_1 e_6+e_6(-2 a_1+e_6)) \\ 
+ I^+(-,+) \, (3 c_1^2+e_1^2-2 e_1 e_6+e_6 (-2 a_1+3e_6))
\end{dmath*} 
\end{dgroup*} 
Here, $I^\mp_\Lambda(-, -)$ are {\itshape intraband contributions} and $I^\mp_\Lambda(-, +)$ are {\itshape interband contributions} to the particle-particle loop $L^-$ and the particle-hole loop $L^+$, respectively, which are defined explicitly as
\begin{align} \label{int_I}
 & I^\mp_\Lambda(\ell_1, \ell_2) = \\[5pt] \nonumber
& \frac{\sqrt 3}{4\pi} \h\int_0^\infty \! \de \kappa \, \kappa \, \dot \chi_\Lambda(e_{\ell_1}\mh(\kappa )) \, \chi_\Lambda(e_{\ell_2}\mh(\kappa )) \, F^\mp(e_{\ell_1}(\kappa ), e_{\ell_2}(\kappa )) \,,
\end{align}
where $e_\ell(\vec k) = E_\ell(\vec k) - \mu$ are the eigenvalues of the free Hamiltonian measured relative to the chemical potential, $\chi_\Lambda$ denotes the regulator function, and the functions $F^\mp$ are conventionally defined as in \cite{Schober}.

\begin{figure}[t]
\subfloat[Intraband contribution $I_{\Lambda}^-(-1, -1)$ to $L_{\Lambda}^-$.\label{fig_pp_intra}]{\includegraphics[width=0.48\columnwidth]{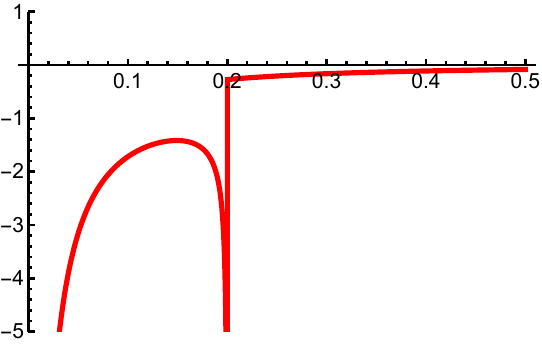}}
\hfill
 \subfloat[Interband contribution $I_{\Lambda}^-(-1, +1)$ to $L_{\Lambda}^-$. \label{fig_pp_inter}]{ \includegraphics[width=0.48\columnwidth]{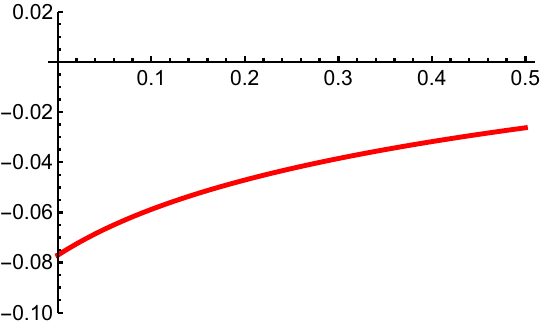}} \\
\subfloat[Intraband contribution $I_{\Lambda}^+(-1, -1)$ to $L_{\Lambda}^+$.
 \label{fig_ph_intra}]{\includegraphics[width=0.48\columnwidth]{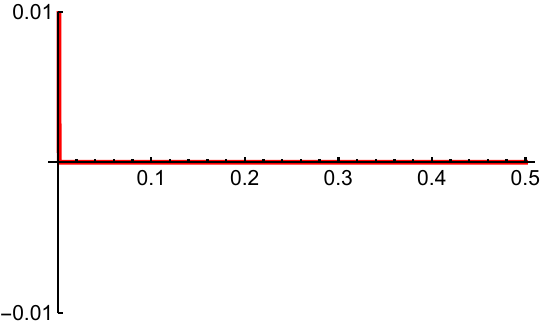}}
\hfill
 \subfloat[Interband contribution $I_{\Lambda}^+(-1, +1)$ to $L_{\Lambda}^+$.
 \label{fig_ph_inter}]{\includegraphics[width=0.48\columnwidth]{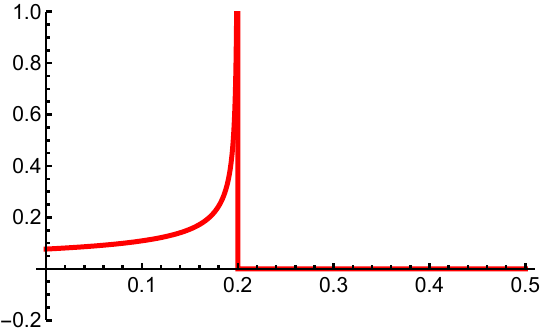}}
 \caption{Particle-particle loop $L_{\Lambda}^-$ and particle-hole loop $L_{\Lambda}^+$ as functions of the scale $\Lambda$.}
 \label{fig_pp_ph}
 \end{figure}

We approximate the free Hamiltonian near each time-reversal invariant momentum by the Rashba Hamiltonian \eqref{RH} with $\gamma = 1$ and $\alpha = 2$, such that the energy bands are given by $ E_{\mp}(\vec \kappa) = \lvert \vec \kappa \rvert ^2 \, \mp \, 2 \lvert \vec \kappa \rvert$. We choose the chemical potential
$ \mu = -0.8$, which lies slightly above the bottom of the lower band (at $E_{\rm min} = -1.0$), and a tiny temperature corresponding to $\beta = 10^4$. Furthermore, we use  a strict regulator function,
\begin{align}
 \chi_\Lambda(e) = \varTheta(\lvert e \rvert -\Lambda) = \varTheta(e-\Lambda) + \varTheta(-e-\Lambda) \,.
\end{align}
For these parameters, Figs.~\ref{fig_pp_intra} and \ref{fig_pp_inter} show the intraband contribution $I_{\Lambda}^{-}(-1, -1)$ and the interband contribution $I_{\Lambda}^{-}(-1, +1)$ to the particle-particle loop $L_\Lambda^{-}$ as functions of the scale $\Lambda$. The interband contribution is small compared to the intraband contribution and approaches a finite value for $\Lambda \to 0$. By contrast, the intraband contribution diverges for $\Lambda \to 0$ and is singular at $\Lambda = 0.2$, which corresponds to the difference between the chemical potential and the minimum of the lower band. 
Figs.~\ref{fig_ph_intra} and \ref{fig_ph_inter} show the the intraband contribution $I_{\Lambda}^{+}(-1, -1)$ and the interband contribution $I_{\Lambda}^{+}(-1, +1)$ to the particle-hole loop $L_\Lambda^{+}$ as functions of the scale $\Lambda$. In this case, the intraband contribution nearly vanishes for all $\Lambda > 0$, while the interband contribution is singular at $\Lambda = 0.2$ and approaches a finite value for $\Lambda \to 0$.

\begin{figure}[t]
\subfloat[$U_1 > 0$: original couplings]{\includegraphics[width=0.48\columnwidth]{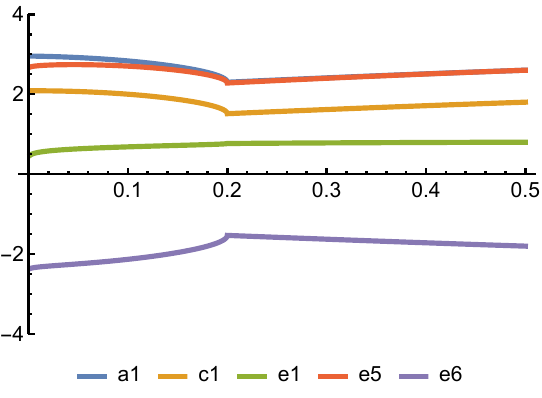} \label{fig_res_4}} \hfill
\subfloat[$U_1 < 0$: original couplings \label{subfig:original_couplings}]{\includegraphics[width=0.48\columnwidth]{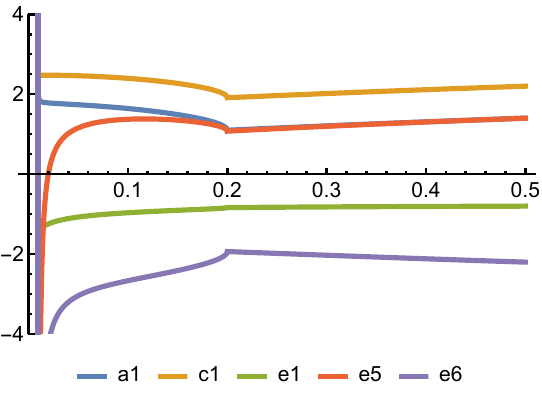} \label{fig_res_1}} \\
\subfloat[$U_1 < 0$: transformed couplings \label{subfig:transformed_couplings}]{\includegraphics[width=0.48\columnwidth]{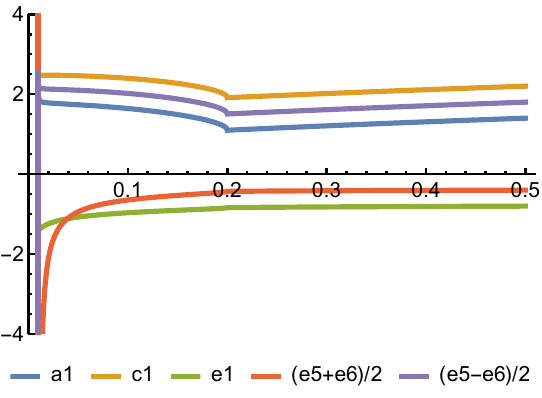} \label{fig_res_2}} \hfill
\subfloat[$U_1 < 0$: absolute values of transformed couplings \label{subfig:results}]{\includegraphics[width=0.48\columnwidth]{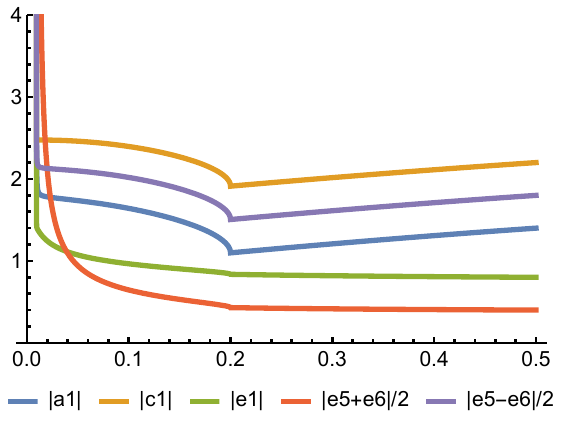} \label{fig_res_3}}
\caption{Numerical solution of the g'ology equations with repulsive onsite interaction $U_0 = 2$ and nearest-neighbor interaction $U_1 = \pm 0.2$.}
\label{fig:results}
\end{figure}

Next, Fig.~\ref{fig:results} shows our numerical solution of the g'ology equations for a repulsive onsite interaction, $U_0 = 2$, and a nearest-neighbor interaction, $U_1 = \pm 0.2$ (at the initial scale $\Lambda_0 =0.5$), which employs the above results for the loop terms.
For $U_1 < 0$, we find  a divergence of the flow at a critical scale $\Lambda^*$ (see Fig.~\ref{fig_res_1}). However, the same divergence is not obtained for $U_1 > 0$ (see Fig.~\ref{fig_res_4}). To analyze this divergence, Fig.~\ref{fig_res_2} shows the flow of linearly transformed couplings, and Fig.~\ref{fig_res_3} their absolute values. It is clearly seen that the coupling $(e_5 + e_6)$ becomes dominant in the flow and triggers the divergence.
From this, one can show that the effective interaction kernel at the critical scale is given by
\begin{align} \label{sup_int}
& V_{s_1 s_2 s_3 s_4}(X_1, X_2, X_3, X_4) = \\[5pt] \nonumber
& v_* \h (\delta_{X_1 X_3} \delta_{X_2 X_4} - \delta_{X_1 X_4} \delta_{X_2 X_3} ) \, [\sigma_x]_{s_1 s_2} \h [\sigma_x]_{s_3 s_4} \,,
\end{align}
where  $v_* < 0$. 

The above result for the effective interaction serves as the starting-point of our mean-field analyis. We consider the superconducting interaction operator
\begin{align} 
 \hat V & = -\frac 1 2 \, \int \! \db^2 \vec k \int \! \db^2 \vec k' \!\sum_{X_1, \ldots, X_4} \sum_{s_1, \ldots, s_4} \! V_{X_1 s_1, \ldots, X_4 s_4}(\vec k, \vec k') \nonumber \\[3pt]
 & \quad \, \times \hat a_{X_1 s_1}^\dagger(-\vec k) \, \hat a_{X_2 s_2}^\dagger(\vec k) \, \hat a_{X_3 s_3}(\vec k') \, \hat a_{X_4 s_4}(-\vec k') \,, \label{gen_int_sup}
\end{align}
where we have abbreviated $\hat a_s^{(\dagger)}(X + \vec k) \equiv \hat a_{X s}^{(\dagger)}(\vec k)$ and used the normalized measure
\begin{equation} \label{nmeasure}
 \int \! \db^2 \vec k = \frac{1}{|\mathcal B|} \h \int_{\mathcal B} \de^2 \vec k \,,
\end{equation}
where $\lvert \mathcal B \rvert$ denotes the surface area of the Brillouin zone. The {\itshape superconducting interaction kernel}
\begin{align} \label{kernels}
V_{X_1 s_1, \ldots, X_4 s_4}(\vec k, \vec k') = V_{s_1 s_2 s_3 s_4}(X_1, X_2, X_3, X_4) 
\end{align}
is independent of $\vec k$ and $\vec k'$ and given by Eq.~\eqref{sup_int}. The {\itshape mean-field interaction} is then obtained by the standard substitution of $\hat a_{X_3 s_3}(\vec k') \h \hat a_{X_4 s_4}(-\vec k')$ by a thermal expectation value with respect to the mean-field Hamiltonian itself (see e.g.~\cite[Eq.~(189)]{Schober}). Thus, we obtain 
\begin{align}
\hat V_{\mathrm{mf}} & = \frac 1 2 \, \int \! \db^2 \vec k \sum_{X_1, X_2} \sum_{s_1, s_2} \bar \Delta_{X_1 s_1, X_2 s_2}(\vec k) \nonumber \\[3pt]
& \quad \, \times \hat a_{X_1 s_1}^\dagger(\vec k) \, \hat a_{X_2 s_2}^\dagger(-\vec k) + \mathrm{H.a.} \,,
\end{align}
where the $(8 \times 8)$ {\itshape gap function} is $\vec k$-independent and factorizes into a spatial part and a spinorial part:
\begin{align} \label{gap_parts}
\bar \Delta_{X_1 s_1, X_2 s_2}(\vec k) = \Delta_{X_1 X_2} \h[\sigma_x]_{s_1 s_2} \,,
\end{align}
with the spatial part given by
\begin{align} \label{gap_spatial}
& \Delta_{X_1 X_2} = \\[5pt] \nonumber
&  -2\hh v_* \int \! \db^2 \vec k' \sum_{s_3, s_4} [\sigma_x]_{s_3 s_4} \, \big\langle \hat a_{X_1 s_3}(\vec k') \, \hat a_{X_2 s_4}(-\vec k') \big\rangle\,.
\end{align}
The anticommutation relations of fermionic annihilation operators imply the antisymmetry of its spatial part,
\begin{align}
\Delta_{X_1 X_2} = - \Delta_{X_2 X_1} \,,
\end{align}
such that the diagonal entries vanish,
\begin{align}
\Delta_{OO} = \Delta_{AA} = \Delta_{BB} = \Delta_{CC} = 0 \,.
\end{align}
By the $C^3$ symmetry, there are only two independent {\itshape gap parameters,}
\begin{align} \label{def_param_1}
\frac 1 {\sqrt 3} \, \Delta_0 & :=  \Delta_{OA} =  \Delta_{OB} =  \Delta_{OC} \,, \\[3pt]
\frac 1 {\sqrt 3} \, \Delta_1 & :=  \Delta_{AB} =  \Delta_{BC} =  \Delta_{CA} \,. \label{def_param_2}
\end{align}
Hence, the spatial part of the gap function is given by
\begin{align} \label{gap_spatial_mat}
\Delta = \frac 1 {\sqrt 3} \left( \begin{array}{rrrr}
0 & \Delta_0 & \Delta_0 & \Delta_0 \\
-\Delta_0 & 0 & \Delta_1 & -\Delta_1 \\
-\Delta_0 & -\Delta_1 & 0 & \Delta_1 \\
-\Delta_0 & \Delta_1 & -\Delta_1 & 0 
\end{array} \right).
\end{align}
Thus, we have shown that the gap function is of spin-triplet type. 
This result differs from the simple Rashba model, for which we found in \cite[Eq.~(199)]{Schober} that the gap function is of spin-singlet type.

Next, the {\itshape mean-field Hamiltonian,} $\hat H^{\mathrm{mf}} = \hat H^0 + \hat V^{\mathrm{mf}}$, can be written in matrix form as follows:
\begin{widetext}
\begin{align}
\hat H^{\rm mf} - \mu \hh \hat N &= \frac 1 2 \h \int \! \db^2 \vec k \sum_{X_1, \h X_2} \sum_{s_1, \h s_2} \big( \h \hat a\dag_{X_1 s_1}(\vec k), \, \hat a_{X_1 s_1}(-\vec k) \h \big) \label{eq_mfham} \\[5pt] \nonumber 
 &\quad \, \times \Bigg( \!\! \begin{array}{cc} \delta_{X_1 X_2} \left(H^0_{s_1 s_2}(\vec k) - \mu \h \delta_{s_1 s_2}\right) & \bar \Delta_{X_1 s_1, X_2 s_2}(\vec k) \\[8pt] -\bar \Delta^*_{X_1 s_1, X_2 s_2}(-\vec k) & \delta_{X_1 X_2} \left(- (H^0_{s_1 s_2})^*(-\vec k) + \mu \h \delta_{s_1 s_2} \right) \end{array} \! \Bigg) \, \Bigg( \! \begin{array}{c} \hat a_{X_2 s_2}(\vec k) \\[8pt] \hat a\dag_{X_2 s_2}(-\vec k) \end{array} \! \hspace{-1pt} \Bigg) \,.
\end{align}
\end{widetext}
It can be diagonalized by a {\itshape Bogoliubov transformation} (see e.g. \cite{SigristUeda, Schober}), which in the present case reads
\begin{align}
 \hat a_{Xs}(\vec k) & = \sum_{N} \bar {\mathcal X}_{Xs, N}(\vec k) \, \hat b_{N}(\vec k) \nonumber \\[-3pt]
& \qquad \quad \, + \bar {\mathcal Y}_{Xs, N}(\vec k) \, \hat b\dag_{N}(-\vec k) \,, \label{bogo_1} \\[8pt]
 \hat a\dag_{Xs}(-\vec k) & = \sum_{N} \bar {\mathcal Y}_{Xs, N}^*(-\vec k) \, \hat b_{N}(\vec k) \nonumber \\[-3pt]
& \qquad \quad \, + \bar {\mathcal X}_{Xs, N}^*(-\vec k) \, \hat b\dag_{N}(-\vec k) \,, \label{bogo_2}
\end{align}
where we have introduced the multi-index $N = (n, m, \ell)$ with $n \in \{0, 1\}$, $m \in \{0, 1\}$, and $\ell \in \{-1, +1\}$. With this, the mean-field Hamiltonian is equivalent to
\begin{align}
 \hat H^{\rm mf} - \mu \hat N = \int \! \db^2 \vec k \, \sum_N \varepsilon_N(\vec k) \, \hat b\dag_N (\vec k) \h \hat b_N(\vec k) \,,
\end{align}
where $\varepsilon_N$ are the mean-field energies.
The $(8 \times 8)$ matrices $\bar{\mathcal X}$ and $\bar{\mathcal Y}$ can be calculated analytically by diagonalizing the $(16 \times 16)$ mean-field matrix. As a result, the mean-field energies are given by
\begin{align} \label{eq_mf_ene}
\varepsilon_{n,\h\ell=\mp}(\vec k) =\sqrt{ \left( \frac{\hbar^2 \h \lvert \vec k \rvert ^2 }{2 m^*} - \mu \right)^{\!\!2} + \lvert \Delta_n \rvert ^2} \, \mp \alpha_{\mathrm R} \h\lvert \vec k \rvert \,.
\end{align}
Since they do not depend on the sub-index $m$, each mean-field energy is two-fold degenerate.

The $(8 \times 8)$ {\itshape pairing wave function}, i.e., the order parameter, is defined as
\begin{align} \label{def_order}
\bar\Psi_{X_1 s_1, \h X_2 s_2}(\vec k) = \big\langle \hat a_{X_1 s_1}(\vec k) \, \hat a_{X_2 s_2}(-\vec k) \big\rangle \,.
\end{align}
Explicit calculation shows that it is of spin triplet type,
\begin{align} \label{RGE_op}
\bar\Psi_{X_1 s_1, \h X_2 s_2}(\vec k) = [\Psi_{\mathrm t}]_{X_1 X_2} [\sigma_x]_{s_1 s_2} \,,
\end{align}
where the spin-triplet amplitude is given by
\begin{align} \label{res_psi_triplet}
[\Psi_{\mathrm t}]_{X_1 X_2}(\vec k) & = \sum_{Y} \sum_{nm} \Delta_{X_1 Y} \h \mathcal X_{Y, nm} \\[2pt] \nonumber
& \quad \, \times \frac{1 - f(\varepsilon_{n-}(\vec k)) - f(\varepsilon_{n+}(\vec k))}{2 \h \varepsilon_{n, 0}(\vec k)} \, \mathcal X^*_{X_2, nm} \,,
\end{align}
with the unitary matrix
\begin{align} \label{def_X-matrix}
\mathcal X = \frac 1 {\sqrt 6} \left( \begin{array}{rrrr}
\sqrt 6 & 0 & 0 & 0 \\
0 & -\sqrt 2 & 0 & -2 \\
0 & -\sqrt 2 & \sqrt 3 & 1 \\
0 & -\sqrt 2 & -\sqrt 3 & 1
\end{array} \right).
\end{align}
In the numerator of Eq.~\eqref{res_psi_triplet} the Fermi distribution $f$ is evaluated at the mean-field energies \eqref{eq_mf_ene}, whereas the denominator contains the auxiliary function
\begin{align} \label{def_e0}
\varepsilon_{n, 0}(\vec k) = \sqrt{ \left( \frac{\hbar^2 \, \lvert \vec k \rvert ^2}{2 m^*} - \mu \right)^{\!\!2} +\lvert \Delta_n \rvert ^2} \,.
\end{align}
Using that $\sigma_x = \sigma_z \h \i \sigma_y$, Eq.~\eqref{RGE_op} further implies that the pairing vector points in the positive $z$ direction. These results for the order parameter also differ from the simple Rashba model, where we found in \cite[Sct.~IV.C]{Schober} that the order parameter is of a mixed singlet-triplet type.

Finally, the gap equation can be derived by inserting our result \eqref{RGE_op} for the order parameter into Eq.~\eqref{gap_spatial}. Thereby, we obtain self-consistent equations for $\Delta_0$ and $\Delta_1$, which turn out to be equivalent to
\begin{align} \label{gap_eq_comp}
1 = -2 \hh v_* \int \! \db^2 \vec k \,\h \frac{1 - f(\varepsilon_{n-}(\vec k)) - f(\varepsilon_{n+}(\vec k))}{ \varepsilon_{n, 0}(\vec k) } \,. 
\end{align}
These are two decoupled equations for the two gap parameters $\Delta_0$ and $\Delta_1$ defined in Eqs.~\eqref{def_param_1}--\eqref{def_param_2}. Note that Eq.~\eqref{gap_eq_comp} does not agree with the standard gap equation \cite[Eq.~(241)]{Schober}, in particular because the denominator is not given by the mean-field energies but by the auxiliary function \eqref{def_e0}.
At zero temperature, Eq.~\eqref{gap_eq_comp} is solved by
\begin{align} \label{sol_gap}
\lvert \Delta_0 \rvert = 2 \h E_{\mathrm R} \, \e^{-c_0/ c_1} \, \e^{-1/(2 \hh c_1 v^* \mhh D_0 )} \,,
\end{align}
where $E_{\mathrm R}$ is the Rashba energy, $D_0 = 2 \pi m^*/(\hbar^2 \h \lvert \mathcal B \rvert)$ the density of states of a single band $e(\vec k) = \hbar^2 \h \lvert \vec k \rvert^2 / (2 m^*)$, $\lvert \mathcal B \rvert$ the area of the hexagonal Brillouin zone, and $c_0$ and $c_1$ are positive constants. The same result is obtained for $\Delta_1$.
Again, this is in contrast to the simple Rashba model \cite[Eq.~(266)]{Schober}, in particular because the constant density of states $D_0$ appears in the exponent of Eq.~\eqref{sol_gap} instead of the density of states of the Rashba$^4$ model under consideration.

In conclusion, we have calculated an RG-effective action which results from the interplay of scattering between inequivalent Rashba points in the Brillouin zone of a tight-binding model on the hexagonal two-dimensional lattice. In the case of a nearest-neighbor attraction, we find an instability towards a symmetry-broken state. We use the low-energy effective action determined by the RG flow to study the symmetry-broken phase and find a novel triplet superconducting state. The model considered here is part of a larger class of models with multiple Rashba points, which is discussed in a second paper \cite{SchoSa2}. 

This work is supported by Deutsche Forschungsgemeinschaft (DFG, German Research Foundation) under Germany's Excellence Strategy EXC-2181/1 - 390900948 (the Heidelberg STRUCTURES Cluster of Excellence).

\bibliography{mybib}{}
\bibliographystyle{unsrt}

\end{document}